\documentclass[a4paper,final]{appolb}
\usepackage{epsfig}
\usepackage{amsmath}
\usepackage{amsfonts}
\usepackage{paralist}
\begin{document}

\title{Negative conductance in driven Josephson junctions
\thanks{Presented at the 20th Marian Smoluchowski Symposium on Statistical Physics, Krak\'ow,
September  22-27, 2007}
}
\author{Marcin Kostur, Lukasz Machura, Jerzy {\L}uczka
\address{Institute of Physics, University of Silesia,
40-007 Katowice, Poland}
\and Peter Talkner, Peter H\"anggi
\address{Institut f\"ur Physik, Universit\"at Augsburg,
86135 Augsburg, Germany}
}
\maketitle

\begin{abstract}
  We investigate an optimal regime of negative-valued conductance,
  emerging in a resistively and capacitively shunted Josephson
  junction, which is driven simultaneously by both,  a time-periodic (ac) and a
  constant (dc) current. We analyze the current-voltage
  characteristics in the
  regime of absolute negative conductance.  We additionally explore
  the stability of the negative response with respect to the ac-current frequency.
\end{abstract}

PACS numbers:  05.60.-k, 05.45.-a, 74.25.Fy, 85.25.Cp

\maketitle

\section{Introduction}

The problem of far-from-equilibrium transport in periodic systems
continues to attract considerable attention during the last decade.
Research in this field provides an important contribution to the
foundations of thermodynamics and statistical physics. For example,
the interplay of nonlinearity, dissipation, fluctuations and
external driving in the presence of chaotic dynamics can lead to a
number of unusual scenarios of dynamical behavior of Brownian
particles. The most prominent example is a Brownian motor system
\cite{zapata,HanMar2005}. Despite all the interesting features
revealed in the last years like Brownian ratchet transport
\cite{AstHan2002}, current reversals \cite{KosLuc2001},
noise-induced phase transitions and the general fact that  noise can
play constructive role in many cases \cite{GamHan1998}, there exist
still unexplored areas of dynamical phenomena awaiting to become
disclosed. Such a rich plethora of transport behaviors emerges when
inertia effects start to dominate the transport characteristics
\cite{inertia}. As an example we recall the  prominent phenomenon of
anomalous response due to external driving which can appear in
relatively simple systems such as in an ac-driven and dc-biased
Josephson junction \cite{MacKos2007a}.  For the systems in
thermodynamic equilibrium the input-output relationship is in
accordance with the linear response theory (LRT). Yet there are
circumstances where LRT holds perfectly well also far away from
equilibrium \cite{HT}. A good example is the overdamped classical
\cite{LinKos2001} and quantum \cite{MacKos2006b} transport of
Brownian particles in a washboard potential: the velocity is an
increasing function of the small external static force (positive
mobility). However, this intuitive or 'normal' situation where the
effect follows the cause may change radically when the system is
subjected to several forcing degrees of freedom. There exists a
broad variety of physical systems which can exhibit 'anomalous'
behavior. One of them is already mentioned the ratchet effect, which
is caused by a symmetry breaking and a source of non-equilibrium
forcing. When the static force affects the time-periodically driven
massive Brownian particle moving in spatially periodic structures
\cite{LucEPL}, it can respond with a negative differential mobility
(NDM) \cite{NDM,KosMac2006c} or even with an absolute negative
mobility (ANM) \cite{MacKos2007a}. In terms of an electric
transport, one can observe an absolute negative conductance (ANC)
when upon an increase of the static voltage bias, starting out from
zero, a current is induced in the opposite direction. This situation
was experimentally confirmed in p-modulation-doped multiple
quantum-well structures \cite{hop} and semiconductor superlattices
\cite{keay}. ANC (ANM) was also studied theoretically for
ac-dc-driven tunnelling transport \cite{HarGri1997} and in the
dynamics of cooperative Brownian motors \cite{REI99}, for Brownian
transport in systems of a complex topology \cite{EicRei2002a} and in
some stylized, multi-state models with state-dependent noise
\cite{CleBro2002}, to name but a few.

In this paper, we continue our prior studies of the same system
detailed in Refs.  \cite{MacKos2007a,MacKos2007b,KosMac2008a} and
analyze the optimal regime of negative-valued conductance. In
section 2, we present the details of the model of the resistively
and capacitively shunted Josephson junction. In  section 3, we study
novel aspects  current-voltage characteristics in the optimal regime
while putting special emphasis on the frequency dependence of the
ac-driving source on the anomalous response.

\section{Stewart-McCumber model }

We explain the dynamics of the Josephson junction in terms of the
well known Stewart-McCumber model \cite{kautz} in which the current
 through the junction is a sum of a Josephson supercurrent
characterized by the critical current $I_0$, a normal Ohmic current
characterized by the resistance $R$ and a displacement current
accompanied with a capacitance $C$.  The Johnson noise plays the
role of the thermal equilibrium noise which is associated with the
resistance $R$.  The quasi-classical dynamics of the phase
difference $\phi=\phi(t)$ between the macroscopic wave functions of
the Cooper electrons in both sides of the junction is described by
the following equation \cite{kautz}
\begin{eqnarray} \label{JJ1}
\Big( \frac{\hbar}{2e} \Big)^2 C\:\ddot{\phi} + \Big( \frac{\hbar}{2e} \Big)^2 \frac{1}{R} \dot{\phi}
= - \frac{\hbar}{2e} I_0 \sin (\phi) \nonumber + \\
+ \frac{\hbar}{2e} I_d + \frac{\hbar}{2e}I_a \cos(\Omega t+\phi_0) +
\frac{\hbar}{2e} \sqrt{\frac{2 k_B T}{R}} \:\xi (t).
\end{eqnarray}
The dot denotes differentiation with respect to time $t$, $I_d$ and
$I_a$ are the amplitudes of the applied dc-current  and ac-current,
respectively, $\Omega$ denotes the angular frequency of the ac
driving source and $\phi_0$ defines the initial phase value.  The
parameter $k_B$ is the Boltzmann constant and $T$ stands for
temperature of the system. The ubiquitous thermal equilibrium
fluctuations are modeled by $\delta$-correlated Gaussian white noise
$\xi(t)$ of zero mean and unit intensity.

The dimensionless form of this equation then reads
\cite{MacKos2007a,MacKos2007b,KosMac2008a}:
\begin{equation}
\label{JJ2}
\ddot{x} + {\gamma} \dot{ x} =- 2\pi \sin (2\pi x) + f + a \cos(\omega s+\phi_0)
+ \sqrt{2 \gamma D} \; \Gamma(s).
\end{equation}
Here $x=\phi/2\pi$ and the dot denotes differentiation with respect to
the dimensionless time $s=t/\tau_0=\omega_0 t$ , where the characteristic time 
 $\tau_0 = 1/ \omega_0=2\pi \sqrt{(\hbar C/2e I_0}$  and $\omega_0$ is the plasma frequency \cite{MacKos2007a}.
Other dimensionless parameters after dimensionless-scaling assume
the form: friction coefficient ${\gamma} = \tau_0 / RC$; the
amplitude and the angular frequency of the ac-current are denoted by
$a = 2\pi I_a / I_0$ and
$\omega = \Omega \tau_0$, respectively.  The dimensionless bias load
stands for $f= 2\pi I_d/ I_0$, the rescaled zero-mean Gaussian white
noise $\Gamma(s)$ possesses the auto-correlation function $\langle
\Gamma(s)\Gamma(u)\rangle=\delta(s-u)$, the noise intensity $D = k_B
T / E_J$ and the Josephson coupling energy  is $E_J=(\hbar/2e)\:
I_0$. The actual stationary averaged voltage reads
\begin{equation}
\label{V}
V= \frac{\hbar \omega_0}{2e} \; v, \quad v=\langle \dot{ x} \rangle,
\end{equation}
where the stationary dimensionless averaged voltage is $ v=\langle
\dot{x} \rangle $ and the brackets denote an average over the
initial conditions, over all realizations of the thermal noise and
the long time limit over one cycle of the external ac-driving.

\section{Negative conductance}

It often proves useful  to use the mechanical analog of the system
(\ref{JJ2}), being an inertial Brownian particle which performs a
one-dimensional random motion in the periodic potential $U(x)=-\cos
(2\pi x)$. The phase is the equivalent of the space coordinate of
the Brownian particle and the currents plays the role of driving
forces on the particle \cite{KosMac2006c,MachuraJPC}. Because Eq.
(\ref{JJ2}) is equivalent to a set of three ordinary differential
equations, the phase space of (\ref{JJ2}) is three-dimensional. In
consequence, the deterministic ($D=0$) nonlinear system (\ref{JJ2})
typically may exhibit chaotic properties \cite{kautz}. In general,
the deterministic dynamics ($D = 0$) exhibits a very rich behavior.
One can detect periodic, as well as quasi--periodic and chaotic
solutions for a given set of the system's parameters. Switching on
finite thermal noise one thereby activates the diffusive dynamics
where stochastic escape events among existing attractors become
possible. Moreover, the particle can now visit any part of the phase
space and proceed within some finite time interval by following
closely any existing stable or unstable orbits.

Assuming a zero dc-current ($f=0$), the motion is unbiased and
symmetric  in the ac-amplitude. The stationary averaged velocity (or
voltage) in this case then must be zero. The simplest option to
destroy this symmetry is to apply the dc-bias current, i.e. we set
$f \ne 0$.  
It breaks the reflection symmetry $x \to - x$ of the potential and
in turn allows the averaged velocity (voltage) to assume non-zero
values, which typically assume the same sign as $f$. Any deviation
from this rule is counterintuitive. As we have shown previously, a
Josephson junction exhibits many exciting features, including
absolute negative conductance (ANC) \cite{MacKos2007a,MacKos2007b},
negative differential conductance (NDC), negative-valued nonlinear
conductance (NNC) or the reentrant effect of the negative
conductance \cite{KosMac2008a}.  In mechanical, particle-like motion
terms, it corresponds to a negative mobility of the Brownian
particle.

\begin{figure}[htpb]
 \begin{center}
\includegraphics[angle=0,width=0.7\linewidth ]{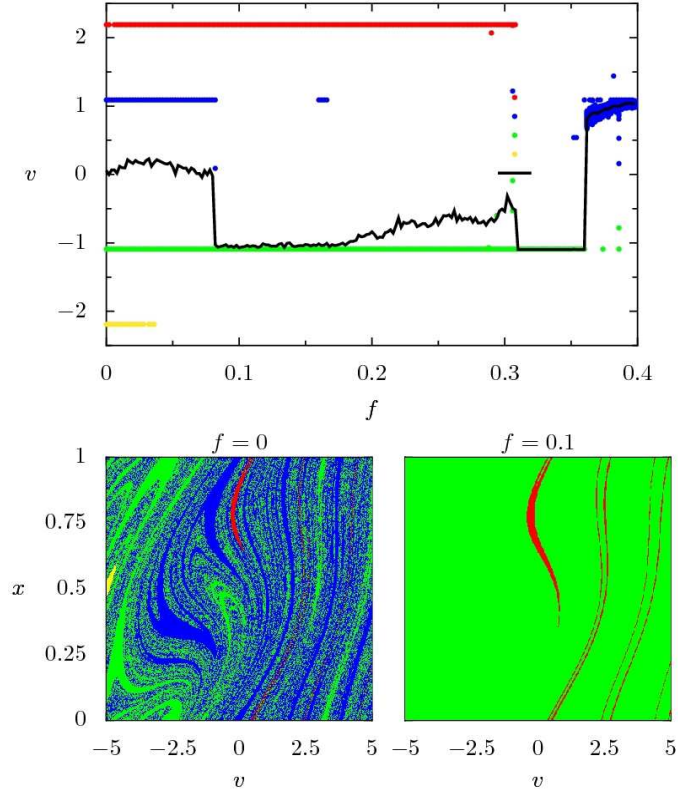}
 \end{center}
\caption{ (color online) Upper panel:  colored lines represent
  the time averaged velocity in the deterministic case $D=0$. The four
  colors correspond to the four attractors $v=-2.2,  -1.1, 1.1,  2.2$.
The black line denotes the long-time velocity (voltage),
  averaged over initial conditions (position, velocity and phase).
  In the
  lower panel, the corresponding basins of attraction are depicted:
  yellow and green mark regimes where  transport occurs in the negative direction (i.e. for
  $v=-2.2$ and $ v=-1.1$, respectively); blue and red mark regimes
  for which the
  attractor is transporting in the positive direction (i.e. for
  $v=1.1$ and $v= 2.2$, respectively).  The averaged velocity (voltage)
 for $f=0.1$ is determined by the structure of the basins of
  attraction rather than by the attractor itself: Although the
  positively transporting attractor (red) possesses a larger velocity  $v$, its basin of
  attraction is much smaller than the one transporting into negative direction.
  Thus, the contribution of this smaller basin
  to the total transport remains small. The presence of
  noise does not markedly change this characteristics (see Fig.  \ref{fig1}
  for $f=0.1$). The remaining parameters read: $a=19.5$, $\gamma = 1.2$,
  $\omega=6.9$. The basins of attraction are shown for the initial
  phase value at $\phi_0= - \pi/2$.  }
\label{fig2}
\end{figure}

\begin{figure}[htpb]
 \begin{center}
\includegraphics[angle=0,width=0.84\linewidth ]{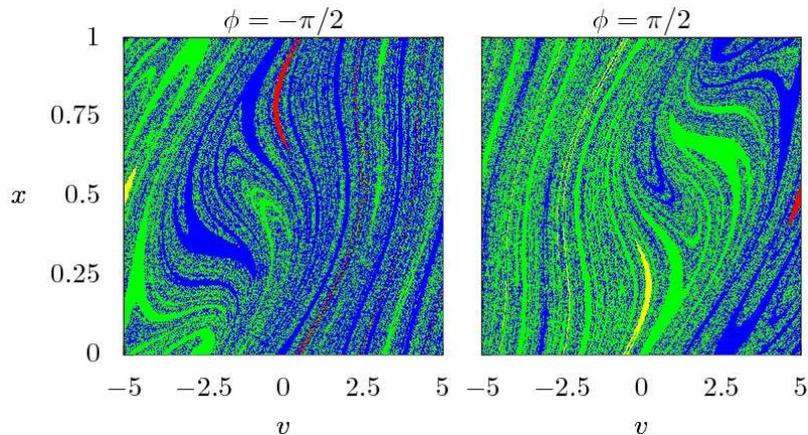}
 \end{center}
\caption{ (color online) An example of two corresponding sets of
  basins of attraction for the deterministic dynamics, $D=0$ at vanishing bias $f=0$.  The
  left part is exactly the same as the left part in the lower panel of  Fig.  \ref{fig2}.  The right part
  corresponds to the transformed initial conditions $(x, v, \phi_0)
  \to (-x, -v, \phi_0 + \pi)$. We recall that the system is periodic with
  period 1.  One can note the present symmetry and consequently it leads to
   zero-voltage after averaging over all initial conditions.  }
\label{fig3}
\end{figure}

In a related study \cite{KosMac2008a}, we found an optimal regime of
the negative conductance in the parameter space $\{a, \omega,
\gamma\}$. This regime is located around the values $a\in(12, 21),
\omega \in (6.5, 7.5)$ and $ \gamma \in (0.9, 1.4)$.  This regime
seems to be optimal in the sense that the negative conductance is
most profound in a relatively large domain with relatively large
values of the dimensionless voltage.
The occurrence of a negative conductance may be governed by various
mechanisms.  In some regimes negative response is induced solely by
thermal equilibrium fluctuations, i.e.  the effect is absent for
vanishing thermal fluctuations. Yet in other regimes, ANC can occur
in the noiseless, deterministic system while a finite  temperature
either destroys the effect or diminishes its strength
\cite{MacKos2007a,KosMac2008a}. Nevertheless, both situations have
its origin in the noise--free ($D=0$) structure of the stable and
unstable orbits. To be more concrete, let us consider a particular set of parameters, namely,  
$a=19.5$, $\gamma = 1.2$  and   $\omega=6.9$.  
For this set of parameters, 
the deterministic behavior can be understood if one studies the
structure of the underlying attractors and the corresponding basins
of attraction.  In the upper panel of Fig.  \ref{fig2}, we show the
long-time averaged voltage. It turns out that for
\newpage
 \noindent   small  bias $f$  
 the system possesses four attractors:  two transporting in the   positive direction, $v=1.1$ and $v=2.2$, and two transporting in the negative
direction, $v=-1.1$ and $v=-2.2$. Notably, if one performs the
average over all initial conditions (position, velocity and initial
phase) with a corresponding, non-weighted uniform distribution then
the resulting voltage $v$ behaves in the way depicted by the black
curve. The zero voltage for the case $f=0$ follows from symmetry
arguments, see in Fig.  \ref{fig3}.
The right panel in this figure can be obtained from the left panel by
the transformation $(x, v, \phi_0) \to (-x, -v, \phi_0+\pi)$ or, what
is easier visible, by the rotation along the origin $(0, 0)$ of the
angle $\pi$.  In other words, for each trajectory transporting to the
right direction there exists its partner transporting exactly in the
opposite direction.  So, both contributions cancel each other and thus
all averages of the voltage over symmetrical distributions in phase
space become zero, yielding $v=0$. When a finite bias is applied, this
symmetry becomes broken, now typically yielding a nonvanishing
velocity $v$.  A small positive bias results in this regime in a
positive voltage (normal response).
A further increase of the dc-current destabilizes two attractors,
namely, $v=-2.2$ and $v=1.1$, and  for example at  $f=0.1$ there
exist only two attractors: $v = -1.1$ and $v = 2.2$, see Fig. \ref{fig2}. By taking the
arithmetic average of these two numbers one could expect that the
voltage $v$ would be positive (and thus no ANC is expected).
However, if one performs the averaging over all initial uniformly
distributed initial conditions then the voltage $v$ is negative and
its value is very close to $v=-1.1$.  To explain this result one has
to inspect the basins of attraction of orbits with $v =-1.1$ and $v
= 2.2$. It turns out that for the positively transporting attractor
$v = 2.2$, the measure of the basin of attraction is much smaller as
compared to the measure of the basin of attraction for the negative
transporting attractor $v = -1.1$. This is depicted with  the right
lower panel of Fig.  \ref{fig2}: The green (light) regions are much
larger than the red (dark) regions.  As a consequence, directed
transport is dominated by orbits which carry a  negative voltage
$v$. 
\begin{figure}[htpb]
 \begin{center}
\includegraphics[angle=0,width=0.84\linewidth ]{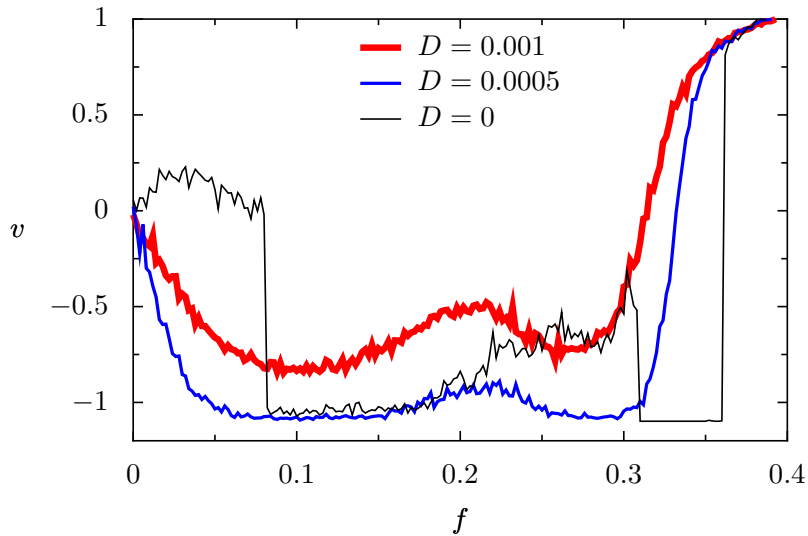}
 \end{center}
\caption{(color online) The dimensionless stationary averaged
voltage $v$ as a
  function of the dc-bias current $f$.  The black (thin) line corresponds to
  the deterministic ($D=0$) dynamics, the blue (intermediate-thick) line
  corresponds to a dimensionless temperature of $D=0.005$, while the red (thick)
  line corresponds to $D=0.001$.  For a non-zero temperature, absolute
  negativer conductance (or absolute mobility) occurs. The remaining  fixed parameters
  read: $a=19.5$, $\gamma = 1.2$, $\omega=6.9$. For small values of
  the dc-current $f<0.05$, the temperature changes dramatically the $v-f$
  characteristics.  }
\label{fig1}
\end{figure}

The influence of temperature is presented in Fig. \ref{fig1},
where we depict the dimensionless stationary
averaged voltage $v$ versus the dimensionless dc-current
$f\in[0,0.4]$ for three selected values of the  noise intensity $D=k_BT/E_J= 0, 0.001, 0.0005$. 
We clearly detect that at non-zero temperature
($D > 0$) and for a small positive dc-current $f$, the voltage $v$
is negative. This case illustrates the phenomenon of the absolute
negative conductance (ANC). Let us emphasize the constructive role
of noise here: In the noiseless, deterministic case ($D=0$) and for
small bias, the system response behaves normal: The averaged voltage
is positive for a positive dc-current (normal transport behavior).
One can  notice   that an increase of the temperature typically
diminishes the effect.
The above ANC effect can be explained  by the fact that the dynamics of the system
located close to the bifurcation point, if perturbed by thermal
noise, takes place in the region of the  phase space where the
stable attractor is emerging beyond the bifurcation point.
\begin{figure}[htb]
 \begin{center}
\includegraphics[angle=0,width=0.84\linewidth ]{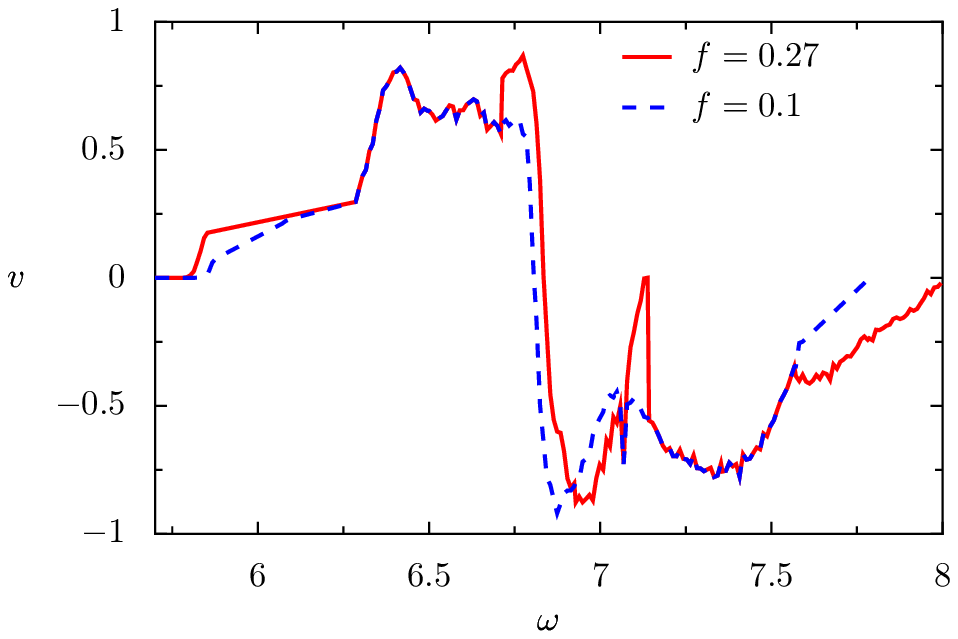}
 \end{center}
\caption{(color online) The stationary, averaged voltage is depicted
{\it vs} the ac-current frequency for $D=0.001$ at two dc-currents
$f=0.1$ and
  $f=0.27$.  The other parameters are as in Fig.  \ref{fig1}.}
\label{fig4}
\end{figure}

Finally, the impact of the frequency of the ac-current on the
average voltage is depicted in Fig.  \ref{fig4} for two different
values of the dc-current: $f=0.1$ (dashed blue line) and $f=0.27$
(solid red line). We see that up to the value of $\omega \simeq 5.8$
the average voltage remains zero.  A further increase of angular
driving frequency leads to finite transport with a positive voltage.
For $\omega \simeq 6.7$ the voltage suddenly drops and crosses over
into a negative average value, thus representing a negative-valued
conductance. This situation remains essentially for the remaining
regime of frequencies up to $\omega \simeq 8$. One can also observe
that in some regions of $\omega$, the voltage stays close to the
same value for both, $f=0.1$ and $f=0.27$, cf. Fig.  \ref{fig4} and
the regime around $\omega \approx 6.5$ or $\omega \approx 7.3$. 
It means that in the current-voltage characteristics one could observe 
Shapiro steps \cite{szapiro}. 
Please note that for both scenarios with $f=0.1$ and $f=0.27$ the
variation of the voltage with increasing angular frequency depicts a
qualitatively robust similar behavior, despite the fact that the
deterministic dynamics can  behave quite different (not shown).

In summary, we put forward an analysis of the negative conductance
occurring in the system of a resistively and capacitively shunted
Josephson junction. For this phenomenon to occur it is necessary
that two driving sources operate simultaneously, namely an ac- and a
dc-current source. We have related the deterministic dynamics with
its stable and unstable orbits to the normal and anomalous response
of the junction to the external stimuli. We are confident that this
very regime of ANC can be successfully tested with an experiment
involving a single Josephson junction, and the presented setup will
be stable within the small variations of any of the structural
parameters.

\section*{Acknowledgments}
\noindent The work supported in part  by the MNiSW Grants N202 13132/3786 (J. {\L}.), 
N202 203534 (M. K. and L. M.) 
and the DAAD-MNiSW program "Dissipative
transport and ordering in complex systems".

\bibliographystyle{aipprocl}

\end{document}